\documentclass[a4paper,11pt]{article}
\usepackage{pos}

\title{Probing light dark matter particles with astrophysical experiments}
\author*[a,b]{Tanmay Kumar Poddar}
\affiliation[a]{Theoretical Physics Division, Physical Research Laboratory, Ahmedabad-380009, India\\
   }

\affiliation[b]{Discipline of Physics, Indian Institute of Technology, Gandhinagar-382355, India\\
}

\emailAdd{tanmay@prl.res.in}
\abstract{The evidence of gravitational wave was first indirectly confirmed by the orbital period loss of Hulse-Taylor binary system which agrees well with the Einstein's general relativistic prediction. The perihelion precession of planets, gravitational light bending and Shapiro time delay are other tests of Einstein's general theory of relativity. However there are small uncertainties in the measurements of those observations from the general relativistic prediction. To account those uncertainties, we propose radiation of ultralight axions and vector gauge boson particles in the context of $U(1)^\prime$ extended beyond standard model scenario. We obtain constraints on ultralight axion parameters (axion mass and decay constant) from the observational uncertainties of orbital period loss of compact binary systems, gravitational light bending, Shapiro time delay and birefringence phenomena. We also obtain the bounds on ultralight $U(1)_{L_\mu-L_\tau}$ gauge bosons from the orbital period loss of compact binary systems. The uncertainties in the perihelion precession of planets also put bounds on the $U(1)_{L_e-L_{\mu,\tau}}$ light gauge bosons. These light particles can be promising candidates of fuzzy dark matter which can be probed from the above precision measurements. }

\FullConference{%
  *** The European Physical Society Conference on High Energy Physics (EPS-HEP2021), ***\\
  *** 26-30 July 2021 ***\\
  *** Online conference, jointly organized by Universität Hamburg and the research center DESY ***
}

\begin{document}
\maketitle

\section{Introduction}
The Standard Model (SM) of particle physics is a very successful theory in a sense that it can describe the interaction between all the fundamental particles in nature to a very good accuracy consistent with the experiments. However, the galactic rotation curve and the Bullet cluster experiment by Chandra X ray observatory confirm the presence of non luminous /dark matter (DM) which cannot be explained by SM. The DM constitutes $25\%$ of the total energy budget of the universe and its constituent is still unknown. To explain such beyond standard model (BSM) phenomenon, one can simply add particle or extend the gauge group with the SM. The weakly interacting massive particle (WIMP) model is one of the popular DM models and it can be a favourable DM candidate. Though, direct detection experiment put strong constraints on WIMP DM. Also it cannot explain the small scale structure problem of the universe. To resolve these problems, physicists consider alternative DM models and one of them is the fuzzy dark matter (FDM) model. In the following, we consider axions or axion like particles (ALPs), light gauge bosons as  FDM candidates and they can form Bose-Einstein condensate. The de Broglie wavelength of the FDM particles is of the size of a dwarf galaxy and it behaves as a wave dark matter. In the following, we have considered some astrophysical experiments to probe such ultralight particles.
% The orbital period loss of the Hulse-Taylor (HT) compact binary system is the first indirect signal of gravitational wave (GW) which agrees well with the Einstein's general relativistic (GR) prediction. However, there is less than one percent uncertainty in the measurement of the observation from the 	GR prediction. There are other tests of Einstein's GR theory like perihelion precession of planets, gravitational light bending and Shapiro time delay which support Einstein's GR theory. The uncertainties in the measurements of those experiments from the GR values allow us to consider radiation of some ultralight particles. In the following sections, we probe ultralight axions and light gauge bosons particles from the above mentioned astrophysical experiments.

%In Section 2, we probe ultralight axions from the orbital period loss of comapact binary systems (neutron star-neutron star (NS-NS), neutron star-white dwarf (NS-WD)). In Section 3, we probe $U(1)_{L_\mu-L_\tau}$ gauge bosons from the orbital period loss of NS-NS and NS-WD binary systems. We also probe $U(1)_{L_e-L_{\mu\tau}}$ gauge bosons from perihelion precession of planets in Section 4. Gravittational light bending and Shapiro time delay can also put bounds on ultralight axions which is discussed in Section 5. We also probe light axions from birefringence phenomena in Section 6. Lastly in Section 7, we discuss our results.
\section{Constraints on ultralight axions from compact bianry systems}
%If two compact objects (neutron stars (NS), white dwarfs (WD)) of masses $m_1$ and $m_2$ are in a binary systems with orbital frequency $\Omega$ and if $D$ is the semi-major axis of the orbit, then the amount of energy loss for an arbitrary Keplerian orbit with eccentricity $e$ is 
%\begin{equation}
%\frac{dE}{dt}=\frac{32G}{5}\Omega^6\Big(\frac{m_1m_2}{m_1+m_2}\Big)^2 D^4(1-e^2)^{-{7/2}}\Big(1+\frac{73}{24}e^2+\frac{37}{96}e^4\Big),
%\label{eq:1}
%\end{equation}
%where $G$ is the Newton's gravitational constant. Hence, we obtain the orbital period loss of the compact binary system as
%\begin{equation}
%\dot{P_b}=6\pi G^{-\frac{3}{2}}(m_1m_2)^{-1}(m_1+m_2)^{-\frac{1}{2}}D^{\frac{5}{2}}\Big(\frac{dE}{dt}\Big)
%\end{equation}
For Hulse-Taylor (HT) binary system, the general relativistic (GR) value of the orbital period loss is $\dot{P_b}_{GR}=(-2.40263\pm0.00005)\times 10^{-12}ss^{-1}$ and its observed value is $\dot{P_b}_{observed}=-2.423(1)\times 10^{-12}ss^{-1}$. The observed value matches in good agreement with the GR prediction which is the indirect evidence of GW. However, there is less than $1\%$ uncertainty in the observation from the GR prediction. To resolve this uncertainty, we assume that ultralight ALPs can radiate from the binary system and contribute to the orbital period loss. The orbital frequency of the HT binary system is $\Omega\sim 10^{-19}\rm{eV}$. Therefore, if particles with this small mass (which is in the ballpark of FDM particles) radiate from the binary systems, then it can contribute to the orbital period loss. If a compact object is immersed in a low mass axionic potential and it interacts with the nucleons of the compact star then it can develop a long range axionic hair outside of that object. In Table \ref{tableI}, we consider four compact binary systems and put bounds on the axion decay constants from observation.
\begin{table}[h]
\centering
\begin{tabular}{ lcc  }
 
 \hline
Compact binary system & $f_a$ (GeV) & $\alpha$\\
 \hline
PSR J0348+0432  & $\lesssim 1.66\times 10^{11}$  & $\lesssim 5.73\times 10^{-10}$ \\
 PSR J0737-3039 & $\lesssim   9.69\times 10^{16}$  &$\lesssim 9.21\times 10^{-3}$ \\
 PSR J1738+0333 & $\lesssim 2.03\times 10^{11}$  & $\lesssim 8.59\times 10^{-10}$\\
PSR B1913+16 & $\lesssim 2.07\times 10^{17}$  & $\lesssim 3.4\times 10^{-2}$ \\
 \hline
\end{tabular}
\caption{\label{tableI} Summary of the upper bounds on the axion decay constant $f_a$ from compact binaries. For all the binaries we assume axion mass $m_a<<10^{-19}$ eV which is constrained from the orbital frequencies of the binaries.}
\end{table}  
In Table \ref{tableI}, $\alpha$ denotes the ratio of axion mediated fifth force to the gravitational force. 
%The relic density of axionic FDM is
%\begin{equation}
%\Omega_{FDM}\sim 0.12\Big(\frac{f_a}{10^{17}\rm{GeV}}\Big)^2\Big(\frac{m_a}{10^{-22}%\rm{eV}}\Big)^\frac{1}{2}.
%\end{equation}
From Table \ref{tableI}, we obtain the stronger bound on $f_a$ is $\lesssim 10^{11}\rm{GeV}$ for ALPs of mass $m_a\ll 10^{-19}\rm{eV}$ which implies if ALPs are FDM, they do not couple with quarks\cite{one} .
\section{Vector gauge boson radiation from compact binary systems in a gauged $L_\mu-L_\tau$ scenario}
Long range $L_\mu-L_\tau$ gauge bosons can radiate from the compact binary systems due to the presence of muons in NS $(N\approx 10^{55})$ because of the large chemical potential of degenerate electrons. However, WDs do not constitute muons. The radiation of ultralight gauge bosons of mass $M_{Z^\prime}\ll 10^{-19}\rm{eV}$ can contribute to the orbital period loss of the binary systems. In Table \ref{tableII} we obtain bounds on the gauge coupling for the above mentioned four compact binary systems from the observations of the orbital period loss. In Figure \ref{fig:axion_profile} (upper panel, left) we show the variation of gauge coupling with the gauge boson mass. PSR J1738+0333 gives the stronger bound on the gauge coupling as $g\leq 4.24\times 10^{-20}$ \cite{two}.
\begin{table}[h]
\centering
\begin{tabular}{ lcc  }
 \hline
Compact binary system \hspace{0.5cm} & $g$(fifth force)\hspace{0.5cm} & $g$(orbital period decay)\\
 \hline
PSR B1913+16  & $\leq 4.99\times 10^{-17}$  & $\leq 2.21\times 10^{-18}$ \\
PSR J0737-3039 & $\leq 4.58\times 10^{-17}$  &$ \leq 2.17\times 10^{-19}$\\
PSR J0348+0432 & $ -$  & $\leq 9.02\times 10^{-20}$ \\
PSR J1738+0333 & $ -$  &$ \leq 4.24\times 10^{-20}$\\
% PSR J1738+0333 & $\lesssim 2.03\times 10^{11}$  & $\lesssim 8.59\times 10^{-10}$\\
%PSR B1913+16 & $\lesssim 2.07\times 10^{17}$  & $\lesssim 3.4\times 10^{-2}$ \\
 \hline
\end{tabular}
\caption{\label{tableII} Summary of the upper bounds on the gauge coupling $g$ from the fifth force and orbital period decay for the four compact binaries. For all the binaries we assume $M_{Z^\prime}<<10^{-19}$ eV which is constrained from the orbital frequencies of the binaries.}
\end{table}
\section{Constraints on long range force from perihelion precession of planets in a gauged $L_e-L_{\mu,\tau}$ scenario}
Due to the presence of electrons, $L_e-L_{\mu,\tau}$ gauge force can mediate between Sun and planets and contribute to the uncertainty in the meauserement of the perihelion precession of planets from the GR prediction ($\mathcal{O}(10^{-3})$ for Mercury). The stronger bound on gauge coupling is obtained from Mars as $g\leq 3.5\times 10^{-25}$. The gauge boson mass is constrained from the distance between Sun and planet. In Figure \ref{fig:axion_profile} (upper panel, right) we show the variation of gauge coupling with the gauge boson mass for all planets upto Saturn \cite{three}. 
\section{Constraints on axionic fuzzy dark matter from light bending and Shapiro time delay}
There are uncertainties in the measurement of the observations from the GR prediction which are $\mathcal{O}(10^{-4})$ for gravitational light bending and $\mathcal{O}(10^{-5})$ for Shapiro time delay. We assume that the long range axion mediated Yukawa type fifth force between the celestial bodies (Earth, Sun) can contribute to the uncertainties in the measurements of those observations. The mass of the ALP is constrained by the distance between Earth and Sun which yields $m_a\lesssim10^{-18}\rm{eV}$. From the observations, we obtain the stronger upper bound on the axion decay constant as $f_a\lesssim 9.85\times 10^{6}\rm{eV}$ from Shapiro delay. This implies, if ALPs are FDM they do not couple with quarks. In Figure \ref{fig:axion_profile} (lower panel, left) we obtain the variation of axion decay constant with its mass \cite{four}. 
\section{Probing the angle of birefringence due to long range axion hair from pulsars}
\begin{figure}
\centering
\includegraphics[width=5cm]{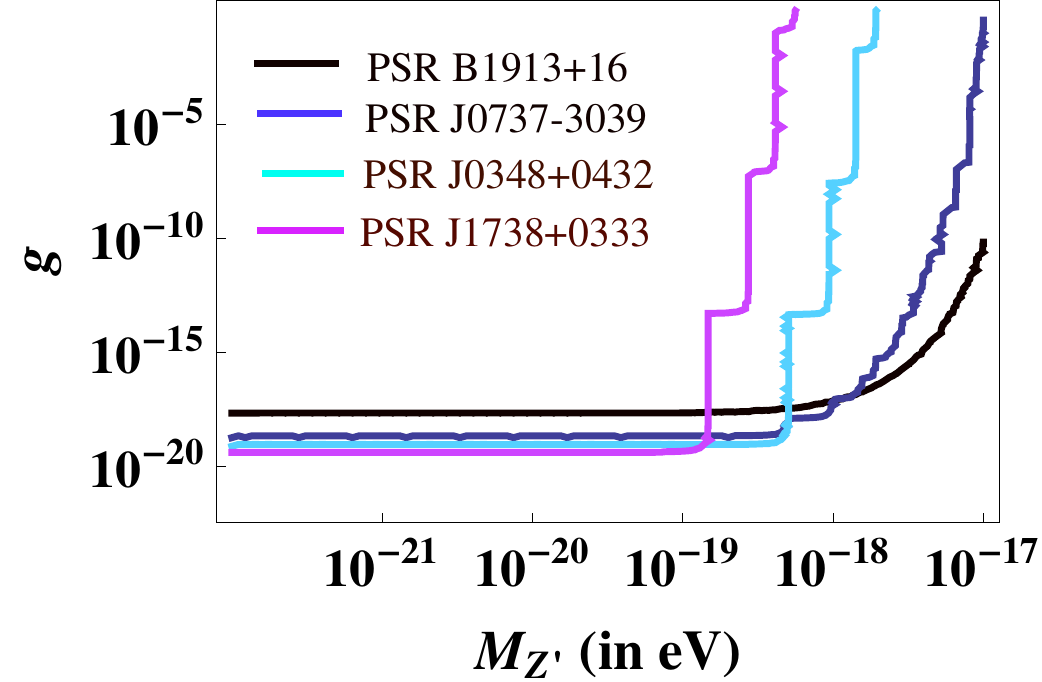}\label{subfig:vvsa}
\includegraphics[width=5cm]{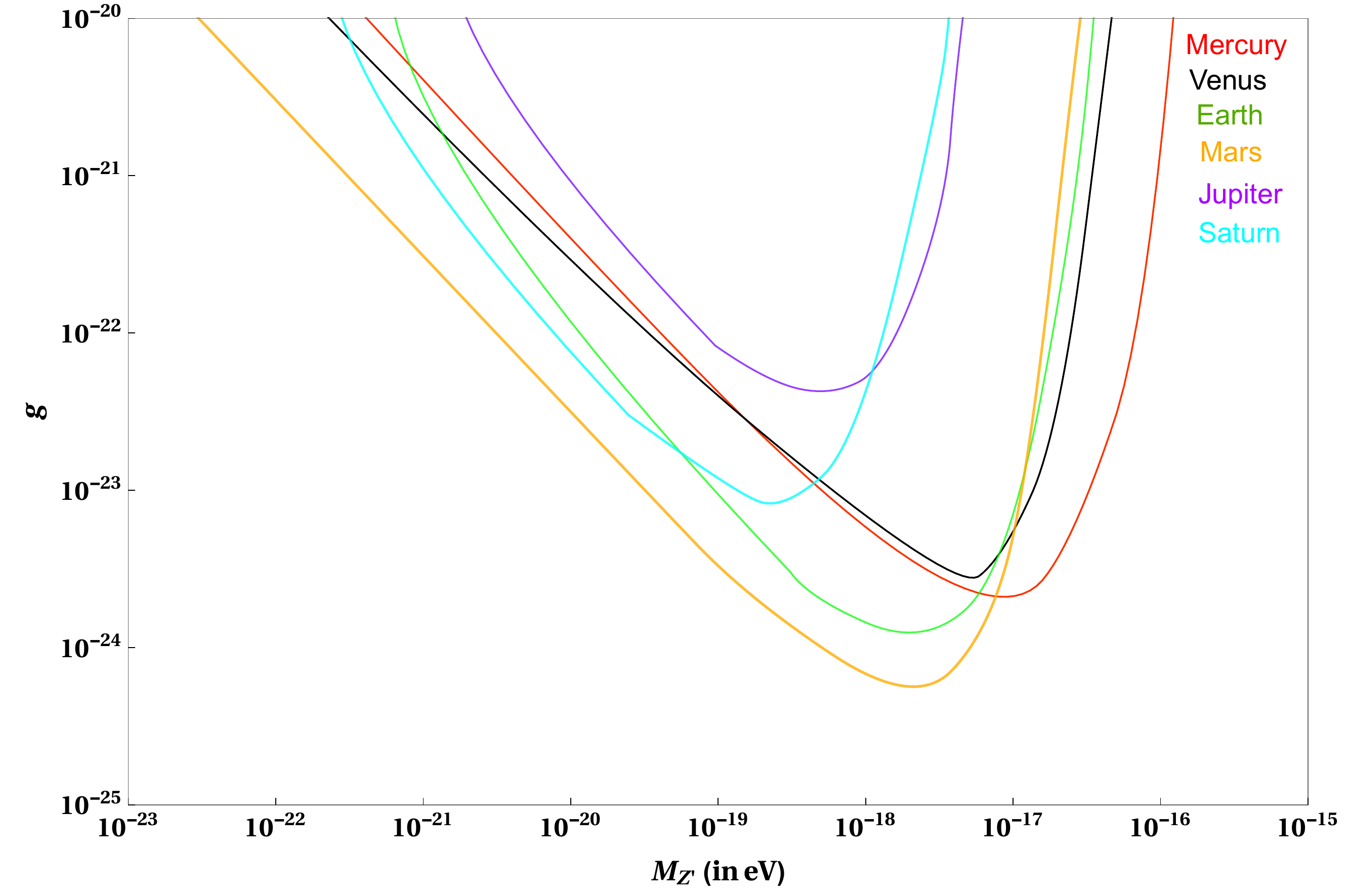}\label{subfig:vvsr}
\includegraphics[width=5cm]{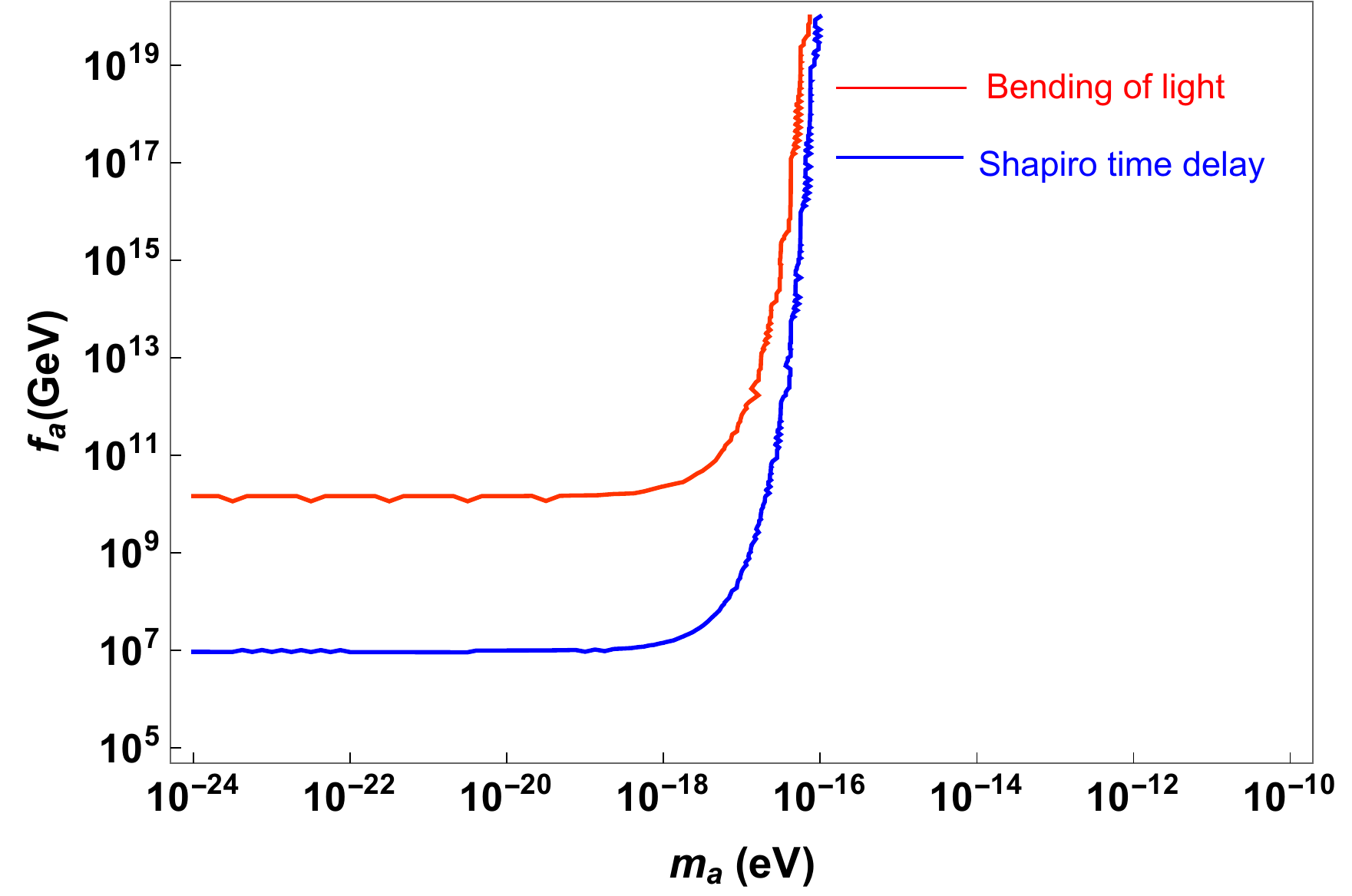}\label{subfig:avsr}
\includegraphics[width=4cm]{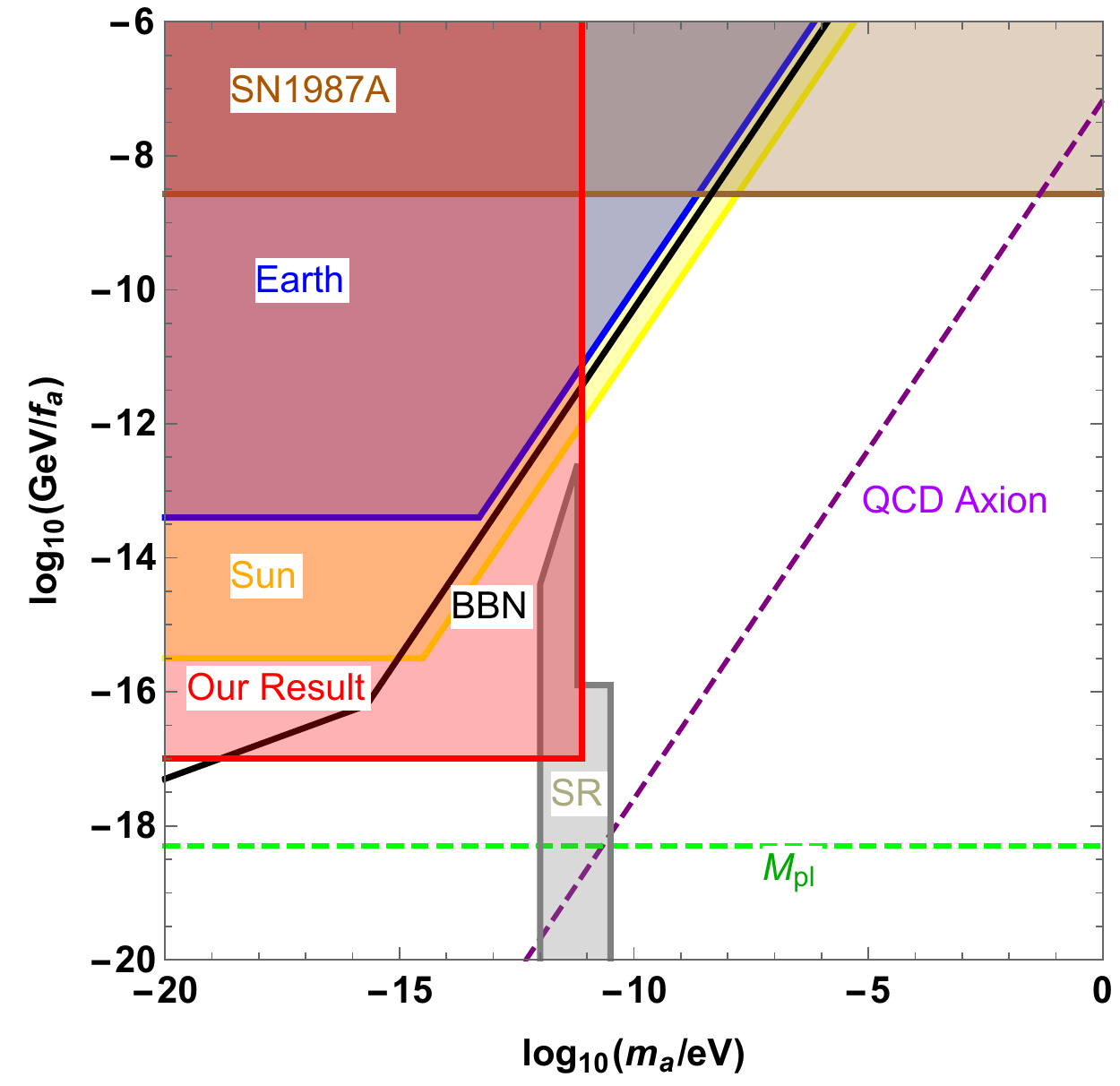}\label{subfig:avsr}
\caption{Variation of gauge coupling of $L_\mu-L_\tau$ type with the mass of gauge boson (upper panel, left). Variation of gauge coupling of $L_e-L_{\mu,\tau}$ type with the gauge boson mass (upper panel, right). Variation of axion decay constant with the axion mass for light bending and Shapiro delay (lower panel, left). Variation of $f_a$ with $m_a$ for birefringence phenomena (lower panel, right).}
\label{fig:axion_profile}
\end{figure}

Rotating neutron star or pulsar can be sourced of long range axions together with the electromagnetic (EM) radiation. When the EM radiation passes through long range axion hair, it can rotate the polarization of the radiation and can produce birefringence. We obtain the birefringent angle due to this axion hair as $0.42^\circ$ which is within the accuracy of measuring the linear polarization angle of pulsar light. Our result continues to hold for axions with $m_a<10^{-11}\rm{eV}$ and $f_a\lesssim 10^{17}\rm{GeV}$ which is shown in Figure \ref{fig:axion_profile} (lower panel, right) \cite{five}.

\end{document}